\documentclass[twocolumn,showpacs,preprintnumbers,amsmath,amssymb]{revtex4}

\usepackage{graphicx}
\usepackage{dcolumn}
\usepackage{bm}

\begin{document}

\setlength{\topmargin}{0.05 in}

\title{An improved RF cavity search for halo axions}

\author{S. J. Asztalos$^a$, R. F. Bradley$^b$, L. Duffy$^c$, C. Hagmann$^a$}
\author{D. Kinion$^a$, D. M. Moltz$^d$, L. J Rosenberg$^a$, P. Sikivie$^c$}
\author{W. Stoeffl$^a$, N. S. Sullivan$^c$, D. B. Tanner$^c$, K. van Bibber$^a$}
\author{D. B. Yu$^e$}

\affiliation{$^a$Lawrence Livermore National Laboratory, Livermore, 
	California 94550}
\affiliation{$^b$National Radio Astronomy Observatory, Charlottesville,
        Virginia 22903}
\affiliation{$^c$Department of Physics, University of Florida, Gainesville,
        Florida 32611}
\affiliation{$^d$Department of Chemistry, University of California, Berkeley,
	California 94720}
\affiliation{$^e$Department of Physics and Laboratory for Nuclear Science,
        Massachusetts Institute of Technology, Cambridge, Massachusetts 02139}

\date{\today}

\begin{abstract}
The axion is a hypothetical elementary particle and cold dark matter 
candidate.  In this RF cavity experiment, halo axions entering a resonant 
cavity immersed in a static magnetic field convert into microwave photons,
with the resulting photons detected by a low-noise receiver.  
The ADMX Collaboration presents new limits on
the axion-to-photon coupling and local axion dark matter halo mass density
from a RF cavity axion search in the axion mass range 1.9--2.3~$\mu$eV, 
broadening the search range to 1.9--3.3~$\mu$eV.
In addition, we report first results from an improved analysis technique.
\end{abstract}


\maketitle

\section{\label{sec:level1}Introduction}
The axion is the pseudo-Goldstone boson \cite{wei1, wil1} implied by the 
Peccei-Quinn solution \cite{pec1} to the
``strong-CP'' problem in QCD (for reviews, see, e.g., \cite{kim1, che1}).  
The axion is also a good cold dark matter candidate \cite{pre1,abb1,din2}, 
and could make a substantial contribution to the
nearby galactic halo mass density, estimated to be approximately 
0.45 GeV/cm$^3$ \cite{gat1}.  Axions are commonly thought to be
thermalized, with energy virial width $O(10^{-6})$ \cite{tur1}, 
and may be detected by the Sikivie RF cavity technique \cite{sik1}.  
This paper describes new results from
an ongoing RF cavity search by the ADMX (Axion Dark Matter eXperiment)
Collaboration.  
We report limits based on predictions for the power 
$P\sim g_{a\gamma\gamma}^2$ deposited in the cavity
from two benchmark axion models, DFSZ \cite{din1,zhi1} and 
KSVZ \cite{kim2, svz1}, where  
$g_{a\gamma\gamma}$ is the effective coupling strength of axions 
to two photons \cite{sik1}.

\section{\label{sec:level1}Experiment}
The experimental apparatus has been described elsewhere \cite{us4}.  
Briefly, halo axions couple to the electric field in a tunable resonant 
cavity plus a DC magnetic field provided by a superconducting solenoid 
surrounding the cavity volume.  The signal is excess power in the cavity 
when the frequency of the TM$_{010}$ mode is close to the energy 
of the halo axions.  The electric field in the cavity is coupled by an
antenna probe to an ultra-low-noise cryogenic preamplifier, followed
by further amplification, mixing, and digitization \cite{us2,us1}.
Figure 1 shows a sketch of the detector and Figure 2 a schematic
diagram of the receiver chain.

\begin{figure}
\includegraphics[angle=0, scale=.18]{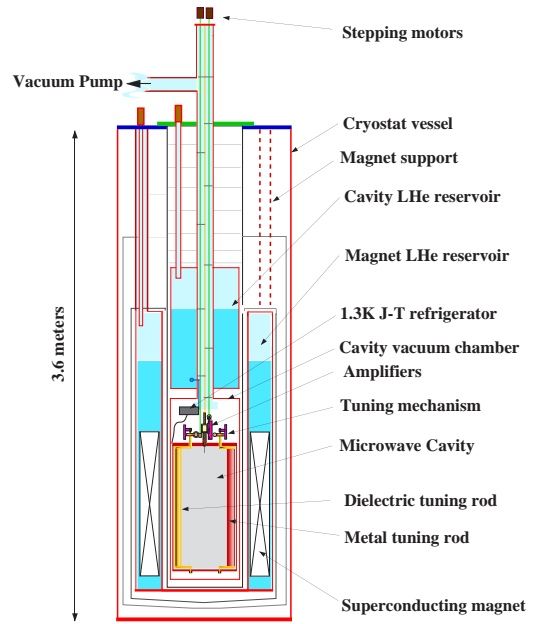}
\caption{Sketch of the RF cavity axion detector.}
\end{figure}

The magnet is a superconducting solenoid of 7.9 T central field.  The
RF cavity is a circular cylinder (50 cm diameter, 100 cm long)
constructed of stainless steel plated with copper and subsequently
annealed.  The unloaded cavity quality factor Q is approximately 200,000
at a resonant frequency of 500 MHz.  In recent running, the temperature 
of the cavity was 1.6 K.  The frequency of the cavity mode is tuned with a pair 
of axial tuning rods (metal or dielectric) that can be translated from 
near the cavity center to the wall.

\begin{figure*}
\includegraphics[angle=0,scale=0.85]{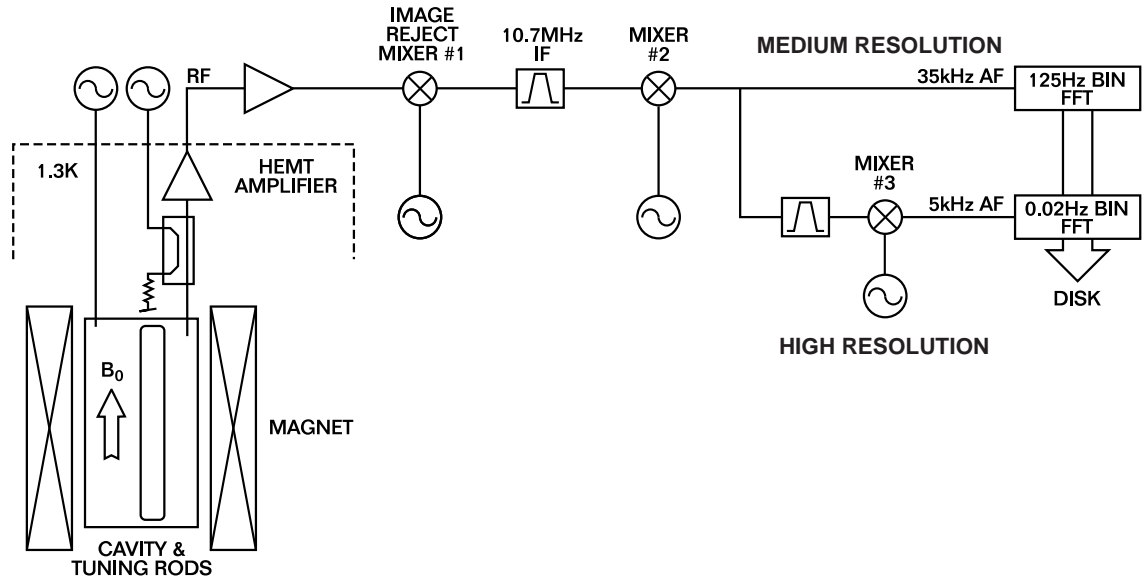}
\caption{Schematic diagram of the receiver chain.}
\end{figure*}

For our most recent run, we used one metal tuning rod ($d \sim 7.5$ cm)
and one ceramic dielectric tuning rod ($d \sim 5.0$ cm), with
the ceramic rod fixed near the cavity center.  With this tuning rod 
configuration, we were able to tune from 461--550 MHz, corresponding 
to axion masses 1.9--2.3~$\mu$eV.

The cryogenic gain of the receiver comes from balanced GaAs HFET preamplifiers
built by the National Radio Astronomy Observatory \cite{daw1}.  At the
operating temperature of 1.6 K, they have a noise temperature 
of approximately 1.9 K and power gain of 17 dB.  The amplifier noise 
temperature was deduced by employing the warmed, critically coupled 
cavity as a Nyquist source.  Due to their improved noise performance
compared to earlier preamplifiers, the system
noise temperature (sum of cavity physical temperature and electronic noise 
temperature) is significantly improved, producing a factor of 2
increase in the search rate relative to earlier operation \cite{us1,us2}.

Data were collected from 08 July 2002 through 27 May 2003 in the form
of ``medium resolution'' single-sided power spectra.  Each spectrum consists of 
10000 averaged, 400-bin, 125 Hz/bin subspectra, with each spectrum having
80 seconds of exposure.  After each 80-second spectrum, the cavity frequency 
was tuned downwards by approximately 1 kHz.  There is also a dedicated
hardware channel optimized for detecting very narrow axion lines; results
from this channel are not reported here.

Each frequency bin appears in 15 to 25 power spectra, corresponding to
an averaging time of about 25 minutes.  The receiver is
stable over a much longer time, as demonstrated in figure 3.  
This figure shows the single-bin rms power divided by the average power 
versus the number of averages (lower abscissa) and the averaging time in
days (upper abscissa).  The log-log slope is approximately $-1/2$ until
$10^{8.5}$ averages, corresponding to 30 days of continuous averaging;
this establishes the ultimate single-bin sensitivity of the receiver, 
corresponding
to $\sim10^{-26}$ Watts, or approximately 1 RF photon per minute at the
signal frequency.

\begin{figure}
\includegraphics[angle=0, scale=.53]{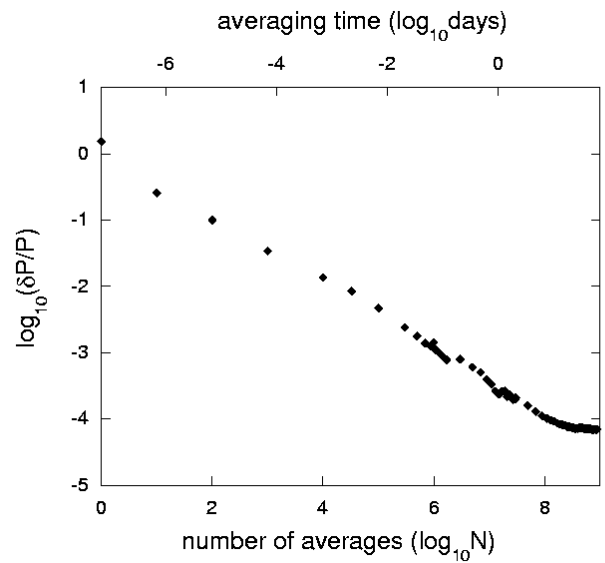}
\caption{Receiver power sensitivity (log$_{10}$ $\delta$P/P) versus number of 
averages (log$_{10}$ N) [lower abscissa] or averaging time (log$_{10}$ days) 
[upper abscissa].}
\end{figure}

\section{Data analysis and Results}
The receiver transfer function was calibrated by recording spectra
with a precision noise source at the receiver input.  Each cavity
spectrum was corrected with this receiver calibration.  The center 175
bins of each spectrum were then corrected with a 5-parameter 
equivalent-circuit model characterizing the cavity, transmission line, 
and amplifier interaction.  Subsequently, spectra were linearly combined 
with a bin-by-bin weighting that accounts for that bin's contribution
to a thermalized axion signal.  More details can be found in \cite{us1,us4}.

This resulting medium resolution combined power spectrum 
was used in two parallel 
analyses.  The first analysis created a power spectrum that was 
the sum of every 6 adjacent bins of the combined power spectrum, 
as was done previously \cite{us1,us4}.  The second analysis, first suggested
in \cite{kra1}, is newly reported here, and applied a Wiener filter (WF) 
\cite{nrc1} derived from the thermalized near-Maxwellian axion signal 
lineshape \cite{tur1}, whose output formed another power spectrum.

The output of the WF is a bin-by-bin sum with weighting 
\begin{equation}
W(\nu) = \frac{S^2(\nu)}{S^2(\nu)+N^2(\nu)}
\end{equation}
where $S(\nu)$ and $N(\nu)$ are the expected signal and noise powers
in each frequency bin of the combined power spectrum 
(by contrast, the first analysis weights 6 adjacent bins with unity
and others zero).  The WF was normalized so as to 
nearly match the number of candidates from the 6-bin search at the 
same candidate threshold.

Figure 4 shows a WF weighting for an axion mass corresponding to 500 MHz.
The abscissa represents the 125 Hz frequency bin offset
from 500 MHz and the ordinate the WF weight.

\begin{figure}
\includegraphics[angle=0, scale=.53]{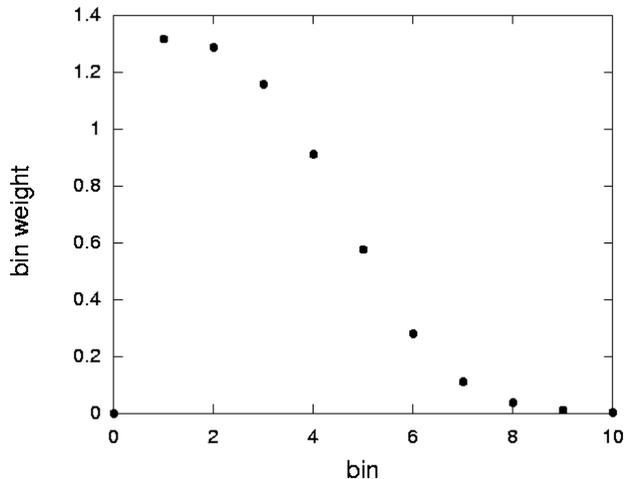}
\caption{Wiener filter weighting for axion mass corresponding to 500 MHz.  
The normalization is discussed in the text.}
\end{figure}

We compared the WF and 6-bin sensitivity by simulating three
different signal lineshapes on Gaussian noise:  the thermalized near-Maxwellian 
lineshape, a narrow single-bin lineshape (developing power in one
bin), and a 10-bin wide lineshape (developing power 
uniformly in 10 adjacent bins).  
For the narrow lineshape, the WF is more sensitive
than the 6-bin filter; this is expected because the WF is narrower.
For the wide lineshape, the WF is slightly less sensitive.  This is 
demonstrated in figure 5, which shows search confidence versus 
candidate threshold in units of single-bin rms power (as in \cite{us1,us4}).  
Sensitivity to narrow lineshapes is important, as narrow structures are 
predicted to accompany galaxy formation \cite{sik2}.

\begin{figure}
\includegraphics[angle=0, scale=0.53]{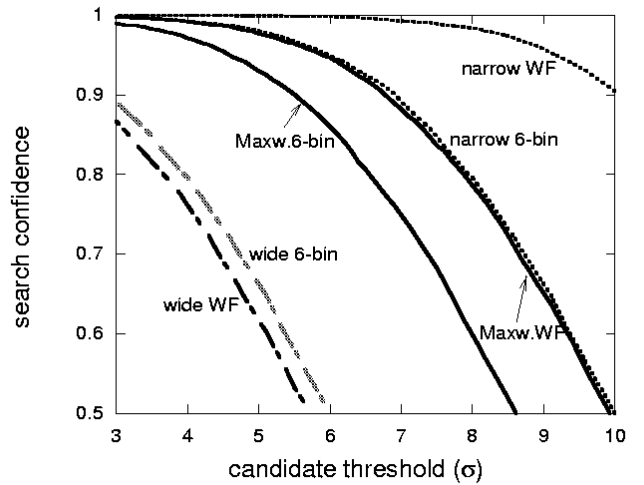}
\caption{Search confidence for finding thermalized axions (solid), 
	narrow 1-bin wide axions (dotted), and wide 
	10-bin wide axions (dashed), from WF and 6-bin searches.}
\end{figure}

Candidates were selected separately from both analyses.
The 6-bin threshold was chosen so as to have the analysis sensitive to 
thermalized KSVZ axions at 94\% confidence, established via Monte 
Carlo technique.  The WF threshold was selected to yield the same or 
slightly fewer number of candidates than the 6-bin search at the same
confidence.  There were 3159 6-bin candidates and
2974 WF candidates (1473 appeared in common), consistent with the
expectation from system noise.

More data were taken at just these candidate frequencies so as to nearly
double the effective integration time.  Of these candidates, 
176 6-bin candidates and 187 WF candidates (76 candidates in common) passed 
a second candidate threshold (chosen to maintain greater than 93\% search
confidence for thermalized KSVZ axions).  Yet more data were taken at 
these surviving candidate frequencies, and candidates again were selected.  
Of these candidates, 12 6-bin candidates and 13 WF candidates (10 candidates 
in common) persisted.  The number of candidates from these successive stages
were approximately consistent with expectations from system noise.

At very high search confidence, ten of the 15 persistent candidates 
did not reappear after integrating for over 2 hours at each candidate 
frequency.  These candidates are therefore highly unlikely to be axions.  
The remaining five candidates (common to both filters) 
reappeared after this long
integration.  These five candidate frequencies were reexamined after removing
a 20 dB attenuator at a calibration port.  The power at each
candidate frequency increased by roughly 100-fold, consistent with the
removal of the attenuator and the hypothesis that these candidates are due to 
environmental RF contamination.  In a further study, the 
cavity was replaced by a stub antenna connected to the input of the receiver.
These same five candidates were also seen in these background power spectra.
The attenuation and stub antenna studies suggest these remaining candidates are 
interfering external signals and are therefore rejected as 
axions.

Figure 6 shows axion-to-photon couplings excluded at greater than 
90\% confidence across our search range.  The upper and lower abscissas
represent axion mass and corresponding microwave photon frequency; the left 
and right ordinates represent power sensitivity in units of expected
thermalized KSVZ axion power, and
the axion-to-photon coupling $g_{a\gamma\gamma}$ divided by axion mass $m_a$. 
The solid line is the 6-bin analysis upper limit, and the 
dotted line is the WF analysis upper limit.
The WF analysis was not applied to frequencies greater than 550 MHz.

\begin{figure}
\includegraphics[angle=0, scale=.45]{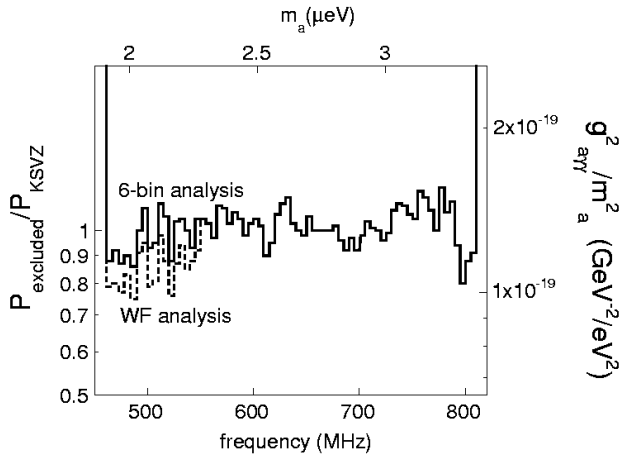}
\caption{Upper limit on axion-to-photon conversion power and coupling 
$g_{a\gamma\gamma}$, excluded at greater than 90\% confidence.}
\end{figure}

These results also constrain the local axion dark matter halo mass density
\cite{us3}.
Figure 7 shows the axion halo mass density excluded
at greater than 90\% confidence for thermalized KSVZ and DFSZ axions.
The upper and lower abscissas represent axion mass and corresponding 
microwave photon frequency; the ordinate represents local axion dark 
matter mass density.  
The lower pair of lines are upper limits for KSVZ axions; the upper 
pair of lines are upper limits for the weaker-coupled 
DFSZ axions.  Solid lines are upper limits obtained from the 6-bin
analysis; dotted lines are the upper limits obtained from the more 
sensitive WF analysis.

\begin{figure}
\includegraphics[angle=0, scale=.53]{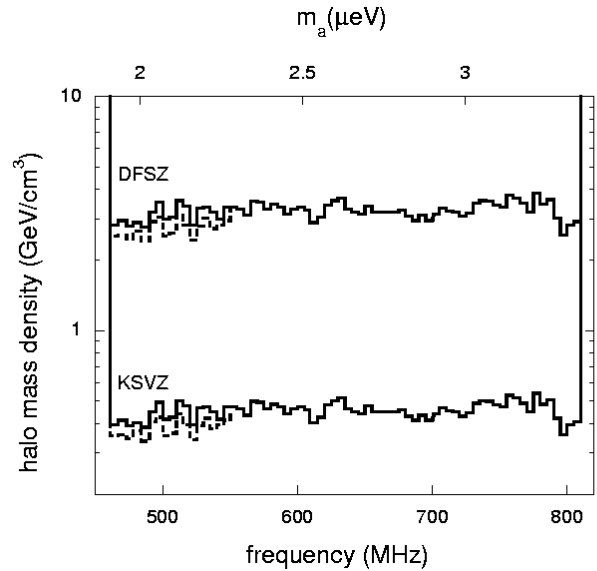}
\caption{Upper limits on galactic axion dark matter halo mass density excluded
at greater than 90\% confidence for KSVZ and DFSZ axions using the two analyses
described in the text.}
\end{figure}

\section{Conclusion}
In conclusion, we now exclude at greater than 90\% confidence a KSVZ halo 
axion of mass 1.9--3.3 $\mu$eV, assuming axions saturate the local dark
matter halo.  We also exclude at greater
than 90\% confidence a local axion dark matter halo mass density of greater than
0.45 GeV/cm$^3$ ($\sim$3 GeV/cm$^3$) for KSVZ (DFSZ) axions.  Improvements to
this experiment come from lower noise preamplifiers; scanning is faster by 
a factor of 2.  The analysis is also improved; the resulting power 
sensitivity gain from the WF is approximately 13\%, representing an 
increased effective integration time of about 25\%.

This work was performed under the auspices of the U.S. Department of 
Energy by the University of California, Lawrence Livermore National 
Laboratory under Contract W-7405-ENG-48, and the University of Florida
under grant DE-FG02-97ER41209.

\bibliography{paper}

\end{document}